\documentstyle[12pt]{article}
\begin{document}

\topmargin 0pt
\oddsidemargin 5mm
\setcounter{page}{1}

{\it Introduction. }--- The concepts Scaling and Universality have played
an essential role in the description of statistical systems \cite{A1}.

 Recently a multisite interaction system on Husimi tree approximation
was investigated \cite{A2}. First, it was shown, that this approach yields
good approximation for the phase diagrams, which closely match the exact
results obtained on a Kagome lattice \cite{A3}. Second, a multisite
antiferromagnetic interaction was studied and interesting connections
with the area of dynamical systems was made. The qualitative picture of
full doubling bifurcations diagram including chaos, period-3 windows,
etc., for the magnetization of the base site of this system was exhibited,
whereas in antiferromagnetic Potts model only one period doubling
occurred \cite{A4}.

 On the other hand, it is well known, that universality of Feigenbaum
constants directly applies to period doubling bifurcations sequence \cite{A5}.

 The aim of our paper is numerical calculation of the Feigenbaum constants
for the three-site antiferromagnetic interaction (TSAI) Ising spin system
and to receive the quantitative description of the transition of this
statistical physical system from ordering to chaos.

{\it Husimi tree and recursion relation. }--- The pure Husimi tree \cite{A6}
is characterized the $\gamma$-the number of the triangle
neighbors.The 0th-generation is a single central triangle.

 The TSAI model in the magnetic field defined by the Hamiltonian
\begin{equation}
\label{R1}
H=-J_3^{'}\sum_{\triangle}{\sigma}_i{\sigma}_j{\sigma}_k-h^{'}\sum_i
{\sigma}_i ,
\end{equation}
where ${\sigma}_i$ takes values $\pm 1$, the first sum goes over all
triangular faces of the Husimi tree and the second over all sites. Besides we
denote $J_3=\beta J_3^{'} ,\ h=\beta h^{'} ,\ \beta=1/kT $,
where h--external magnetic field, T--temperature of the system and $J_3<0$
corresponds to the antiferromagnetic case.

 The partition function will be written as
\begin{equation}
\label{R2}
Z=\sum_{\{\sigma \}} exp\left\{ J_3\sum_{\triangle}{\sigma}_i{\sigma}_j
{\sigma}_k+h\sum_i{\sigma}_i \right\} ,
\end{equation}
where the summation goes over all configurations of system.

 The advantage of the Husimi tree introduced is that for the models
formulated on it, exact recursion relation can be derived. When "cutting
apart" the Husimi tree at the central triangle it separates into 3
identical branches and each of them contains $\gamma-1$ branches. Then
the partition function mey be written
\begin{equation}
\label{R3}
Z=\sum_{\{{\sigma}_0 \}}exp\left\{ J_3\sum_\triangle
{\sigma}_0^{(1)}{\sigma}_0^{(2)}{\sigma}_0^{(3)}+h\sum_j\sigma_0^{(j)}
\right \}{[g_n({\sigma}_0^{(1)})]}^{\gamma -1}
{[g_n({\sigma}_0^{(2)})]}^{\gamma -1}
{[g_n({\sigma}_0^{(3)})]}^{\gamma -1} ,
\end{equation}
where ${\sigma}_0^{(j)}$ are spins of central triangle, n-number of
shells and the equation for one of branches can be written:
\begin{equation}
\label{R4}
g_n({\sigma}_0)=\sum_{\{{\sigma}_i\ne{\sigma}_0\}} exp\left \{ J_3
\sum_{\triangle}{\sigma}_0 {\sigma}_1 {\sigma}_1+h\sum {\sigma}_1+J_3
\sum_{\triangle}{\sigma}_i{\sigma}_j{\sigma}_k+h\sum_i{\sigma}_i \right \} .
\end{equation}
 One of branches, in its turn, can be cut on the site of 1th-generation,
which is the nearest to the central site. Therefor, the expression for
$g_n({\sigma}_0)$ can be rewritten in the form:
\begin{equation}
\label{R5}
g_n({\sigma}_0)=exp\left \{ J_3\sum_{\triangle}{\sigma}_0 {\sigma}_1{\sigma}_1
+h\sum{\sigma}_1 \right \}{[g_{n-1}({\sigma}_1^{(1)})]}^{\gamma-1}
{[g_{n-1}({\sigma}_1^{(2)})]}^{\gamma-1} .
\end{equation}
 From eq.(\ref{R5}) one can easily obtain:
$$g_n(+)=e^{J_3+2h}g_{n-1}^{\gamma -1}(+)g_{n-1}^{\gamma -1}(+)+
2e^{-J_3}g_{n-1}^{\gamma -1}(+)g_{n-1}^{\gamma -1}(-)+e^{J_3-2h}
g_{n-1}^{\gamma -1}(-)g_{n-1}^{\gamma -1}(-) , $$
$$g_n(-)=e^{-J_3+2h}g_{n-1}^{\gamma -1}(+)g_{n-1}^{\gamma -1}(+)+
2e^{J_3}g_{n-1}^{\gamma -1}(+)g_{n-1}^{\gamma -1}(-)+e^{-J_3-2h}
g_{n-1}^{\gamma -1}(-)g_{n-1}^{\gamma -1}(-) . $$
 Let the following variable be introduced:
\begin{equation}
\label{R6}
x_n=\frac{g_n(+)}{g_n(-)} .
\end{equation}
 Then for $x_n$ we can obtain the following recursion relation:
\begin{equation}
\label{R7}
x_n=f(x_{n-1}),\qquad f(x)=\frac{z{\mu}^2x^{2(\gamma-1)}+2\mu x^{\gamma-1}+z}
{{\mu}^2 x^{2(\gamma-1)}+2z\mu x^{\gamma-1}+1} ,
\end{equation}
where $z=e^{2J_3}, \quad \mu=e^{2h}\ and\ 0\le x_n \le1 . $
 The eq.(\ref{R7}) coincides with that obtained by Monroe \cite{A2}, when pair
interaction absents.

For magnetization of the central base site we obtain:
\begin{equation}
\label{R8}
m=\langle \sigma_0 \rangle =\frac {e^hg_n^{\gamma}(+)-e^{-h}g_n^{\gamma}(-)}
{e^hg_n^{\gamma}(+)+e^{-h}g_n^{\gamma}(-)}=\frac{e^hx_n^{\gamma}-1}
{e^hx_n^{\gamma}+1} .
\end{equation}

{\it TSAI system and Feigenbaum constants. }--- As it is mentioned in
Introduction of this paper, the TSAI system is the nonlinear dynamical system
and the qualitative picture of full doubling bifurcations diagrams, chaos
etc., for the magnetization of the base site of it was existed \cite{A2}.

 The questions we want to address in this paper are, how calculate
the constants of Feigenbaum for TSAI system and if calculated
values will coincide with the famous universal Feigenbaum constants:
\begin{equation}
\label{R9}
\alpha=2.500290\dots , \qquad \delta=4.669201\dots .
\end{equation}

Feigenbaum observed for logistic map (see ref.\cite{A7}) two kinds of
scaling: one that the length $2^n$ cycle first appears at a $r_n$ value,
which obeys:
\begin{equation}
\label{R10}
r_n=r_{\infty}-{\it const}\delta^{-n},\qquad n \gg1 ,
\end{equation}
where $r_{\infty}$ the value of r from which the chaotic behavior ensues and
the sequence essentially never repeats itself.

 The other scaling was a special behavior which occurred near the $x^*$
-value for which the map is extremal ($x^*=1/2$ in the logistic map). If
one started out at value for $x^*$ then
\begin{equation}
\label{R11}
-\alpha=\frac{d_n}{d_{n+1}} , \qquad n \gg 1 ,
\end{equation}
where
\begin{equation}
\label{R12}
d_n=f_{R_n}^{2^{n-1}}(x^{*})-x^{*} .
\end{equation}
In eq.(\ref{R12}) $R_n$ are the values of r $(r_1<R_1<\dots<R_n<r_n)$ and
\begin{equation}
\label{R13}
f_{R_n}^{2^n}(x^*)=x^{*} .
\end{equation}
 Note, that values of $R_n$ \quad and \quad $r_n$ have the same scale and
${r_{\infty}}={R_{\infty}}$.
Therefore
\begin{equation}
\label{R14}
R_{\infty}=R_n-{\it const}{\delta}^{-n} .
\end{equation}
 Eqs.(\ref{R11}) and (\ref{R14}) defines two Feigenbaum constants, which
turns out to be "universal".

 Now let us turn to our questions. One can see from eq.(\ref{R7}), that
the role of above mentioned parameter r for TSAI system on Husimi tree
for each fixed temperature plays external magnetic field h. The
recursion function of eq.(\ref{R7}) has one maximum at ${x^*}=1/{\root
{\gamma-1}\of {\mu}}$. Note, that this $x^*$ depend on values of T and
h, whereas in case of logistic map it is a constant. Further, we
numerically solve the eq.(\ref{R13}) and find out the values of $H_n$
($H_n$ is the analog of $R_n$).Using this values of $H_n$ and
eqs.(\ref{R11}) and (\ref{R14}) we calculate the Feigenbaum constants
for TSAI system. All our numerical calculations are done for
${\gamma}=3,\ T=0.3 \quad and \quad J_3=-1$, and are listed in table 1.

 For {\it const.}, presented in eq.(\ref{R14}), which is depend on
family of reflection functions \cite{A8}, for TSAI system we obtain the
following value: ${\it const.}=0.99\dots$, whereas for logistic map it
is 0.12\dots.

 Using the values of $H_n$ and corresponding them $x_1, x_2,\dots x_n$,
we can also calculate the magnetization for base
site (for each cycle of period doubling) of this system by
eq.(\ref{R8}). It means that each $2^n$ period doubling have n values of
magnetization, which should be explained as an arising of a n-sublattice
phase such that $x_1, x_2,\dots x_n$ determine the states on each
sublattice.

 One can see from table 1, that for real statistical physical system obtained
values of $\alpha$ and $\beta$ coincide with famous Feigenbaum constants
(eq.(\ref{R9})) with high accuracy and thereby confirm they universality once
more.

 The some numerical values for magnetization of base site of TSAI system
are listed in table 1. as well.

 It is interesting to note, that if one lets $\gamma=2$ in eq.(\ref{R7})
rather then $\gamma=3$, the above mentioned situation changes
dramatically.

 Let us consider the following system of equations:
\begin{equation}
\label{R15}
\cases{f(x)-x=0\cr f^{'}(x)=-1\cr }
\end{equation}
 The eq.(\ref{R15}) determined the point, where the first doubling bifurcation
is
begun.

 For recursion function  when ${\gamma}=2$, the eq.(\ref{R15}) will have
the form:
\begin{equation}
\label{R16}
\cases{{\mu}^2x^3+z\mu(2-\mu)x^2+(1-2\mu)x-z=0\cr
{\mu}^2x^2-2z{\mu}^2x-(1+2\mu)=0\cr} ,
 \end{equation}
which for any T and h have only nonphysical solutions. Therefore, for
TSAI system when ${\gamma}=2$ the  period doubling bifurcations picture
absents. It means that for this statistical physical system there is not
phase transition of second order when ${\gamma}=2$.

{\it Conclusion. }---  In this paper we have investigated  TSAI Ising  spin
model by  approximating it with Husimi  tree  structures and calculate the
Feigenbaum constants  $\alpha$  and  $\delta  . $
The numerical  results show, that
obtained valued for these constants for real physical system
coincide with the famous universal  Feigenbaum constants with high accuracy.
The quantitative description of the transition from ordering to  chaos is
also obtained. Hence we  see many of the  very intensely studied and by now
familiar properties of dynamical  systems theory,  which gives possibility to
study the statistical physical systems  in a new context and in a simple
manner. In particular, in this paper  with using the well known technique
for dynamical systems, we analitically show, that for  TSAI system  there
is not phase transition of second order when  $\gamma=2$.

We think, that obtained results are interesting and we plan to continue to
investigate this line  of approach for TSAI system and for several  other
systems.

On the other hand, the study of chaotic statistical physical system has opened
new challenges for theories of stochastic processes,  especially in the
direction of stochasticity of vacuum in QCD \cite{A9}. In this direction
the interesting results for Z(Q) gauge  model  on generalized Bethe lattice
was obtained \cite{A10}. It gives bases to suppose  that  TSAI Ising spin
model on Husimi tree approximation can be connected with double plaquette
representation  of the  gauge  theory.

\begin{center}
***
\end{center}

This  work  was partly  supported by the Grant-211-5291 YPI of the German
Bundesministerium  fur  Forshung and Technologie and by the Grant
INTAS-93-633.  he authors are grateful to Prof. Flume,  N. Izmailian,
A. Akheyan, V. Gurzadyan and A. Shahverdyan for useful discussions.

\newpage
{\bf Table 1.}
\begin{center}
\begin{tabular}{|l|l|l|l|l|l|}
\hline
\ \ period & \ \ \ \ \ \ $H_n $ & \ \ \ \ \ \ \ \ $\alpha $
& \ \ \ \ \ \ \ \ \ $\delta $ & magnetization & \ \ \ \ $x_n$\\
doubling & \  & \  & \  & \ \ \ \ \ \ \ \ m & \ \\
\hline
$2^1=2$ & $0.18354515\dots $ & \
 & \  & -0.6782977 & 0.5423560\\
\  & \  & \  & \  & 0.09151673 & 0.999999\\
\hline
$2^2=4$ & $0.28692571\dots$ & $4.86428158\dots$
& $3.50752342\dots $ & \  & \ \\
\hline
$2^3=8$ & $0.31861247\dots$ & $2.19505287\dots$
& $4.32097441\dots$ & -0.8924331 & 0.3457500\\
\  & \  & \  & \  & -0.1373881 & 0.8200590\\
\  & \  & \  & \  & -0.9581313 & 0.2495890\\
\  & \  & \  & \  & 0.1182513 & 0.9733620\\
\  & \  & \  & \  & -0.8506408 & 0.3886090\\
\  & \  & \  & \  &  -0.2286943 & 0.7699640\\
\  & \  & \  & \  & -0.9640809 & 0.2369180\\
\  & \  & \  & \  &  0.1579719 &  0.9999\\
\hline
$2^4=16$ & $0.32607381\dots$ & $2.76000232\dots$ & $4.5870529\dots$ & \
& \  \\
\hline
$2^5=32$ & $0.32770666\dots$ & $2.42990139\dots$ & $4.65118547\dots$ & \
& \  \\
\hline
$2^6=64$ & $0.32805801\dots$ & $2.5381186\dots$ & 4.66503325 & \  & \  \\
\hline
$2^7=128$ & $0.32813334\dots$ & $2.4897987\dots$ & $4.66830065\dots$ & \
& \  \\
\hline
$2^8=256$ & $0.32814947\dots$ & $2.5099532\dots$ &\  & \  & \  \\
\hline
............. & ............. & ............. & ............. & .............
& .............\\
\hline
$2^{\infty}=\infty$ & $0.3281538\dots$ &\  &\  & \  & \  \\
\hline
\end{tabular}
\end{center}

\end{document}